\documentclass[aps,prb,twocolumn,floatfix,showpacs,superscriptaddress]{revtex4-1}
\usepackage{graphicx}
\usepackage{amsmath}
\usepackage{txfonts}
\usepackage[colorlinks,citecolor=magenta,linkcolor=blue]{hyperref}

\begin{document}


\title{Detecting phase transitions and crossovers in Hubbard models using the fidelity susceptibility}
\author{Li Huang}
\affiliation{Science and Technology on Surface Physics and Chemistry Laboratory, P.O. Box 9-35, Jiangyou 621908, China}
\author{Yilin Wang}
\affiliation{Beijing National Laboratory for Condensed Matter Physics, and Institute of Physics, Chinese Academy of Sciences, Beijing 100190, China}
\author{Lei Wang}
\affiliation{Beijing National Laboratory for Condensed Matter Physics, and Institute of Physics, Chinese Academy of Sciences, Beijing 100190, China}
\affiliation{Theoretische Physik, ETH Zurich, 8093 Zurich, Switzerland}
\author{Philipp Werner}
\affiliation{Department of Physics, University of Fribourg, 1700 Fribourg, Switzerland}
\date{\today}


\begin{abstract}
A generalized version of the fidelity susceptibility of single-band and multi-orbital Hubbard models is systematically studied using single-site dynamical mean-field theory in combination with a hybridization expansion continuous-time quantum Monte Carlo impurity solver. We find that the fidelity susceptibility is extremely sensitive to changes in the state of the system. It can be used as a numerically inexpensive tool to detect and characterize a broad range of phase transitions and crossovers in Hubbard models, including (orbital-selective) Mott metal-insulator transitions, high-spin to low-spin transitions, Fermi-liquid to non-Fermi-liquid crossovers, and spin-freezing crossovers. 
\end{abstract}


\maketitle

\section{Introduction\label{sec:intro}} 

Hubbard models play a central role in the theoretical analysis of correlation effects in solids, such as high-temperature superconductivity in cuprates~\cite{RevModPhys.78.17} and unconventional superconductivity in iron-based materials~\cite{RevModPhys.83.1589}. Due to screening, the non-local matrix elements of the Coulomb interaction are suppressed and one can thus hope to qualitatively capture the properties of correlated materials by treating only the on-site interactions. Of particular interest are the phase diagrams of Hubbard models, which already in the single-band case, and even more so in the multi-orbital versions, exhibit a variety of phases with and without long-range order. Even though the exact phase diagrams in high dimensions ($d \ge 2$) are not known yet, numerous methods have been developed to detect and characterize the different phase transitions and crossovers, including the Mott metal-insulator transitions~\cite{RevModPhys.40.677,RevModPhys.70.1039}, high-spin to low-spin transitions~\cite{PhysRevB.85.245110,PhysRevLett.99.126405,kunes}, Landau Fermi-liquid to non-Fermi-liquid crossovers~\cite{PhysRevLett.101.166405,PhysRevB.81.054513}, and spin-freezing crossovers~\cite{PhysRevLett.101.166405,PhysRevB.79.115119}, just to name a few. Identifying these transitions and understanding the underlying mechanisms is an important aspect of modern condensed matter physics. 

The dynamical mean-field theory (DMFT)~\cite{RevModPhys.68.13,RevModPhys.78.865}, which maps a general lattice model onto a quantum impurity model and solves the effective impurity model self-consistently, is probably the most powerful established method to study the phase transitions and crossovers in high-dimensional Hubbard models. Typical criteria for the Mott metal-insulator transitions are the suppression of the quasi-particle weight $Z$ or the spectral weight at the Fermi level $A(\omega = 0)$~\cite{RevModPhys.40.677,RevModPhys.70.1039}. As for the high-spin to low-spin transitions and spin-freezing crossovers, the criteria could be jumps in the local magnetic moment or characteristic changes in the long-time decay of the spin-spin correlation function $\langle S_z(\tau)S_{z}(0)\rangle$~\cite{PhysRevLett.99.126405,kunes,PhysRevLett.101.166405}. In the framework of DMFT, due to the constraints posed by the available quantum impurity solvers~\cite{RevModPhys.68.13,RevModPhys.78.865}, it is generally not a computationally easy task to evaluate the above quantities. For single-site DMFT, the hybridization expansion continuous-time quantum Monte Carlo algorithm (dubbed CT-HYB) is the most widely used and efficient quantum impurity solver~\cite{PhysRevLett.97.076405,PhysRevB.74.155107,PhysRevB.75.155113,RevModPhys.83.349}. Since it is typically implemented on the imaginary-time axis, we have to perform tedious and numerically ill-defined analytical continuations of the Matsubara self-energy function $\Sigma(i\omega)$~\cite{pade1977} and imaginary-time Green's function $G(\tau)$~\cite{jarrell} in order to obtain reliable $Z$ and $A(\omega = 0)$. Unfortunately, the Monte Carlo data for $\Sigma(i\omega)$ and $G(\tau)$ are usually noisy~\cite{PhysRevB.84.075145,PhysRevB.85.205106,PhysRevB.76.205120}, so that a substantial amount of computer time is required for an accurate estimation of $Z$ and $A(\omega)$. In multi-orbital Hubbard models with rotationally invariant interaction, which are common in realistic simulations of materials, it is also numerically expensive to measure the spin-spin correlation function~\cite{PhysRevB.75.155113,PhysRevB.92.155102}. Thus, for systematic scans of phase diagrams, it would be very helpful to establish an easy-to-compute observable which allows detecting (most of) the transitions and crossovers in Hubbard models.
 
In the present work, we show that the fidelity susceptibility could be an observable with the desired properties. Given a Hamiltonian $\hat{H}(\lambda)$, which depends on the parameter $\lambda$, the quantum fidelity $F(\lambda_1, \lambda_2)$ measures the overlap between the two ground state wave-functions $|\Psi_0(\lambda = \lambda_1)\rangle$ and $|\Psi_0(\lambda = \lambda_2)\rangle$. Then the fidelity susceptibility $\chi_{\text{FS}}(\lambda)$ is defined as the second derivative of $\ln{F}$ with respect to the change of $\lambda$~\cite{PhysRevLett.99.095701,PhysRevLett.105.117203,PhysRevE.76.022101}:
\begin{equation}
\chi_{\text{FS}}(\lambda) = -\left.\frac{\partial^2 \ln F(\lambda,\lambda + \epsilon)}{\partial\epsilon^2}\right|_{\epsilon = 0}.
\label{eq:FS}
\end{equation}
We note that the fidelity susceptibility is an important and fundamental concept in quantum information theory, and has a wide range of applications in quantum many-body systems. Gu \emph{et al.}~\cite{PhysRevE.76.022101,gu2010} demonstrated that it exhibits a maximum or even diverges at a quantum critical point and thus provides a convenient probe of quantum phase transitions. Very recently, Wang \emph{et al.} proposed a generic and efficient approach to measure the fidelity susceptibility of correlated fermions, bosons, and quantum spin systems with Monte Carlo sampling~\cite{PhysRevX.5.031007}. They successfully applied this approach to identify crossovers and quantum phase transitions in one- and two-impurity Anderson models~\cite{PhysRevLett.115.236601}. Inspired by these promising developments, we systematically study the behavior of the fidelity susceptibilies of single-band and multi-orbital Hubbard models  in the framework of single-site DMFT. The purpose of the present work is to explore whether and to what extent we can use the fidelity susceptibility to probe and characterize various phase transitions and crossovers in Hubbard models~\cite{RevModPhys.40.677,RevModPhys.70.1039,PhysRevB.85.245110,PhysRevLett.99.126405,kunes,PhysRevLett.101.166405,PhysRevB.79.115119,PhysRevB.81.054513}.

The rest of this paper is organized as follows: Section~\ref{sec:method} defines the models used in this study and describes the Monte Carlo estimator for the measurement of the (orbital-resolved) fidelity susceptibility. The results are presented in Sec.~\ref{sec:results}, where we demonstrate how the fidelity susceptibility may be used to identity various phase transitions and crossovers. Finally, a summary and discussion are given in Sec.~\ref{sec:con}. Besides the fidelity susceptibility, some related observables may be used to detect the transitions and crossovers. They are discussed in the Appendix.   

\section{Formalism\label{sec:method}}
\subsection{Models\label{subsec:model}} 

In the present study, we limit our discussion to Hubbard models 
\begin{equation}
\hat{H} = -t \sum_{\langle ij \rangle, \sigma}c^{\dagger}_{i\sigma}c_{j\sigma} + \sum_i \hat{H}^{i}_{\text{loc}},
\end{equation}
where $\hat{H}^{i}_{\text{loc}}$ is the local Hamiltonian on each site $i$. In the case of the single-band Hubbard model, $\hat{H}_{\text{loc}}$ reads (omitting the site index $i$)
\begin{equation}
\label{eq:1band}
\hat{H}_{\text{loc}} =  -\mu \sum_{\sigma}n_{\sigma} + U n_{\uparrow} n_{\downarrow}.
\end{equation}
In multi-orbital Hubbard models with Slater-Kanamori type interaction, $\hat{H}_{\text{loc}}$ has the form 
\begin{align}
\label{eq:2band}
\hat{H}_{\text{loc}} = &- \mu \sum_{\alpha\sigma}n_{\alpha\sigma} + U\sum_{\alpha} n_{\alpha\uparrow}n_{\alpha\downarrow} \\
    & + U^{\prime} \sum_{\alpha > \gamma, \sigma} n_{\alpha\sigma}n_{\gamma\bar{\sigma}}
    + (U^{\prime} - J) \sum_{\alpha > \gamma, \sigma} n_{\alpha\sigma}n_{\gamma\sigma} \nonumber \\
    & - J \sum_{\alpha \neq \gamma} (d^{\dagger}_{\alpha\downarrow}d^{\dagger}_{\gamma\uparrow}d_{\gamma\downarrow}d_{\alpha\uparrow} + d^{\dagger}_{\gamma\uparrow}d^{\dagger}_{\gamma\downarrow}d_{\alpha\uparrow}d_{\alpha\downarrow} + h.c.).\nonumber
\end{align} 
Here $\alpha$ and $\gamma$ are the orbital indices, $\sigma=\{\uparrow, \downarrow\}$ the spin index, $\mu$ the chemical potential, $U$ ($U^{\prime}$) the intra-orbital (inter-orbital) Coulomb interaction, and $J$ the Hund's exchange interaction. Unless otherwise specified, $\mu$ is chosen to satisfy the half-filling condition. The $U$ ($U^{\prime}$) and $J$ parameters fulfill the relation $U^{\prime} = U - 2J$ to respect the rotational invariance of the Coulomb interaction. In this paper, a semicircular density of states with half bandwidth $D = 2t$ is used, which corresponds to the infinite-dimensional Bethe lattice. We solve these Hubbard models using single-site DMFT~\cite{RevModPhys.68.13,RevModPhys.78.865} with a state-of-the-art CT-HYB quantum Monte Carlo impurity solver~\cite{RevModPhys.83.349,PhysRevB.74.155107,PhysRevLett.97.076405,PhysRevB.75.155113}. 

\subsection{Generalized fidelity susceptibility\label{subsec:fidel}} 
In the context of DMFT studies, one can choose the tuning parameter $\lambda$ in Eq.~(\ref{eq:FS}) as the hybridization strength between the local impurity and the bath. Hence, the fidelity susceptibility quantifies the sensitivity of the system's state with respect to a variation in the hybridization strength, which differs drastically between different phases. This motivates us to use the fidelity susceptibility as a general tool to detect phase transitions and characterize the different phases and correlation regimes in the Hubbard model. 

The exact Monte Carlo estimator for the impurity fidelity susceptibility in the CT-HYB algorithm reads~\cite{PhysRevLett.115.236601,PhysRevX.5.031007} 
\begin{equation}
\label{eq:fs_wang}
\chi_{\text{FS}}(\lambda) = \frac{\langle \kappa_{\text{L}} \kappa_{\text{R}} \rangle - \langle \kappa_{\text{L}} \rangle  \langle \kappa_{\text{R}} \rangle}{2\lambda^2},
\end{equation}
where $\kappa_{\text{L}}$ and $\kappa_{\text{R}}$ count the number of impurity electron operators ($d^{\dagger}$ or $d$) located in the range $[0,\beta/2)$ and $[\beta/2,\beta)$ of the imaginary-time axis, respectively. Here $\beta=1/(k_{\text{B}}T)$ is the inverse temperature. However, Eq.~(\ref{eq:fs_wang}) cannot be directly applied to lattice models, because $\lambda$ only appears in the auxiliary quantum impurity model and thus typically changes during the DMFT self-consistent iterations~\cite{RevModPhys.68.13,RevModPhys.78.865}. In addition, this estimator is orbital-independent which limits its application to the multi-orbital Hubbard models. Therefore, we ignore the denominator in Eq.~(\ref{eq:fs_wang}) and consider the orbital-dependent correlation function 
\begin{equation}
\tilde{\chi}_{\text{FS}}^{\alpha\gamma} = \langle \kappa^{\alpha}_{\text{L}} \kappa^{\gamma}_{\text{R}} \rangle - \langle \kappa^{\alpha}_{\text{L}} \rangle \langle \kappa^{\gamma}_{\text{R}} \rangle.
\label{eq:chitilde}
\end{equation}
We use the tilde symbol to distinguish this generalized fidelity susceptibility from the original one. The CT-HYB algorithm maps the quantum impurity model to a statistical mechanics problem, i.e. randomly distributed hybridization events on the imaginary-time interval. Quantum phase transitions or crossovers of the quantum impurity model manifest themselves as changes in the distributions of these hybridization events~\cite{PhysRevLett.115.236601}. Equation~(\ref{eq:chitilde}) computes the covariance of hybridization events and is sensitive to various phase transitions of the quantum impurity model. Moreover, resolving the orbital indices provides additional information about the local physics of the quantum impurity. 

In the following, we study the generalized fidelity susceptibility $\tilde{\chi}_{\text{FS}}^{\alpha\gamma}$ as a function of various physical parameters, such as the interaction strength, the chemical potential, etc. With several representative examples we will show that it also captures the critical fluctuations associated with a generic phase transition irrespective of the details of the system, which makes it a very useful and versatile tool for detecting diverse phase transitions and crossovers in the Hubbard models.

\section{Results\label{sec:results}}
\subsection{Mott metal-insulator transition\label{subsec:mit}} 

\begin{figure}[t]
\centering
\includegraphics[width=\columnwidth]{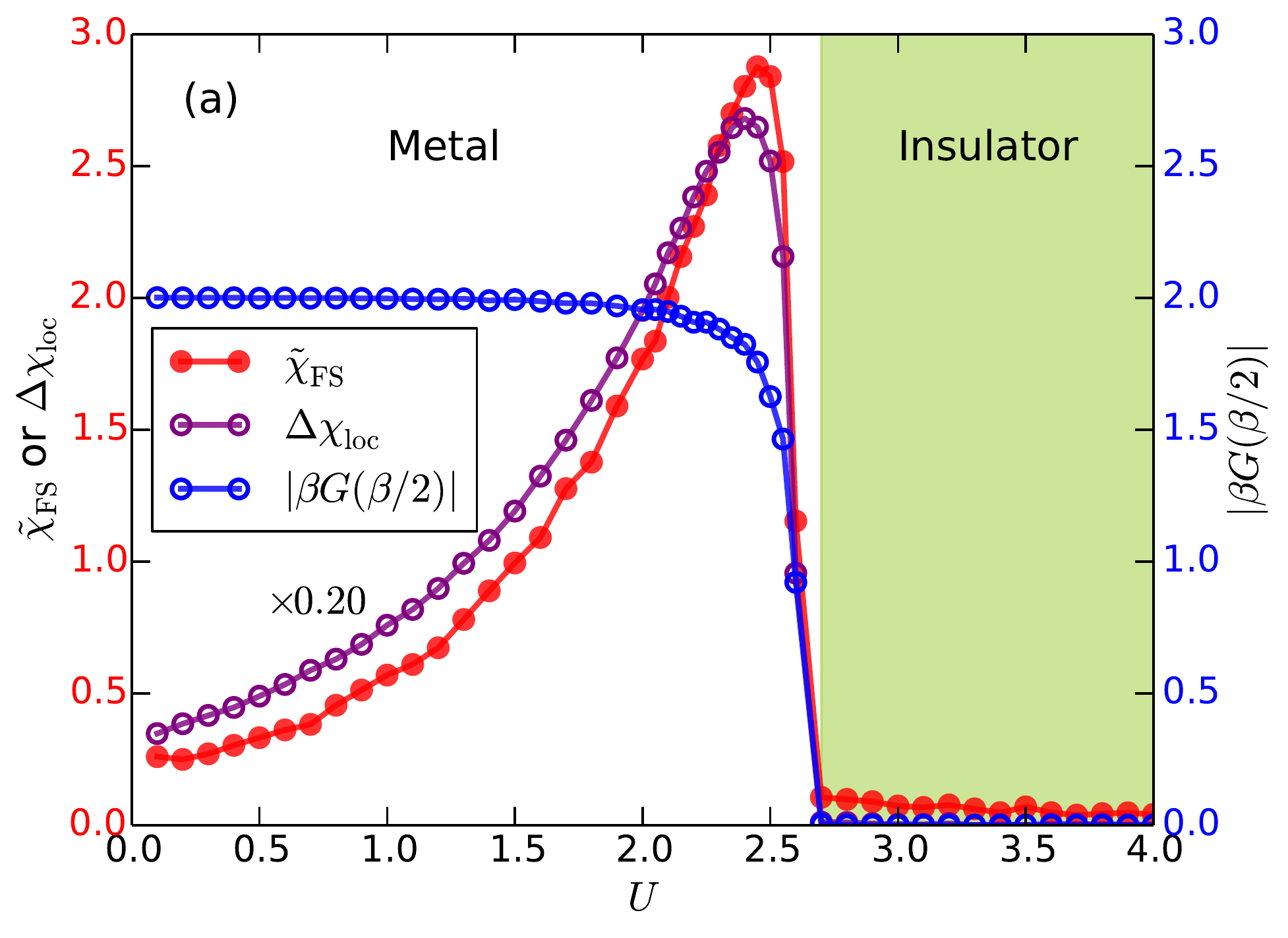}
\includegraphics[width=\columnwidth]{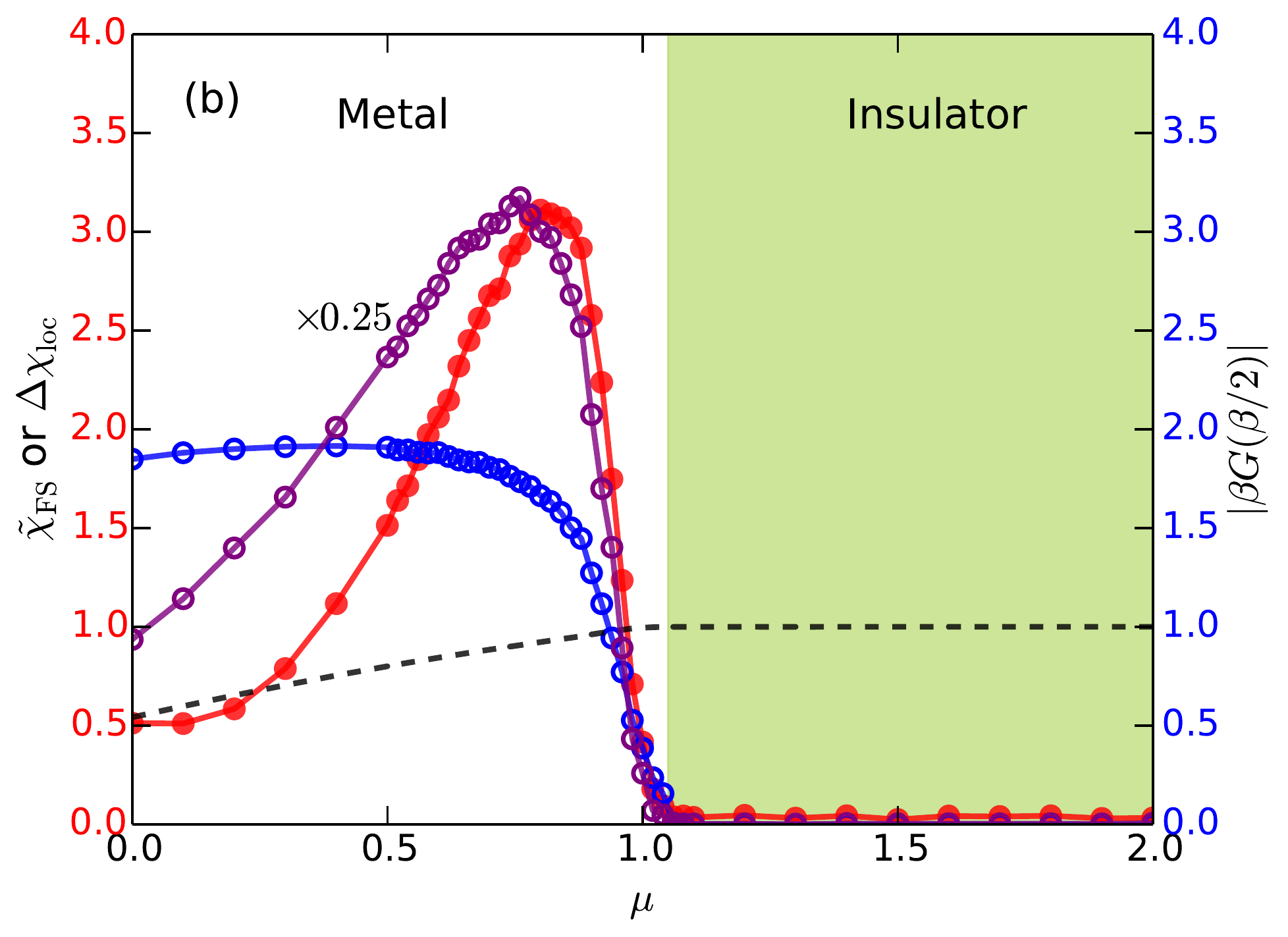}
\caption{(Color online) Mott metal-insulator transition in the single-band Hubbard model on a Bethe lattice ($\beta = 100.0$, $t = 0.5$). (a) Interaction-driven transition at half-filling, i.e., $\mu = U/2.0$. (b) Doping-driven transition for $U = 4.0$. The definition for $\Delta \chi_{\text{loc}}$ can be found in Eq.~(\ref{eq:cloc}), and the growth of this quantity indicates the emergence of local magnetic moments (See Sec.~\ref{subsec:spin} for more explanations). The dashed line shows the total occupation number. The green region indicates the Mott insulating phase. \label{fig:1band}}
\end{figure}

First, we focus on the simplest case, the Mott metal-insulator transition in the single-band Hubbard model, and consider two different scenarios: interaction-driven and doping-driven transitions. The calculated $\tilde{\chi}_{\text{FS}}$ as a function of $U$ and $\mu$ are shown in Fig.~\ref{fig:1band}(a) and (b), respectively. Besides $Z$ and $A(\omega = 0)$, the observable $|\beta G(\beta/2)|$ is often used to identify the metal-insulator transition. It is proportional to $A(\omega = 0)$ at low temperature~\cite{EPL.84.37009,PhysRevB.86.075130},
\begin{equation}
A(\omega = 0) = \frac{1}{\pi} \lim_{\beta \rightarrow \infty} |\beta G(\beta/2)|.
\end{equation}
For the purpose of comparison we also plot this quantity in Fig.~\ref{fig:1band}. We find that $\tilde{\chi}_{\text{FS}}$ exhibits a finite value in the metallic state, and rapidly drops to a tiny value near the metal-insulator transition. The critical points $U_c$ and $\mu_c$ determined from the $\tilde{\chi}_{\text{FS}}$ curves are consistent with those determined by $|\beta G(\beta/2)|$. Hence, $\tilde{\chi}_{\text{FS}}$ is a reliable tool to detect the Mott metal-insulator transition. Note that the obtained $\tilde{\chi}_{\text{FS}}$ shows prominent peaks near the Mott transitions, which are related to the appearance of local magnetic moments near the Mott phase. We will discuss this issue in more detail below.   

\subsection{Orbital-selective Mott transition\label{subsec:osmt}}
 
\begin{figure}[t]
\centering
\includegraphics[width=\columnwidth]{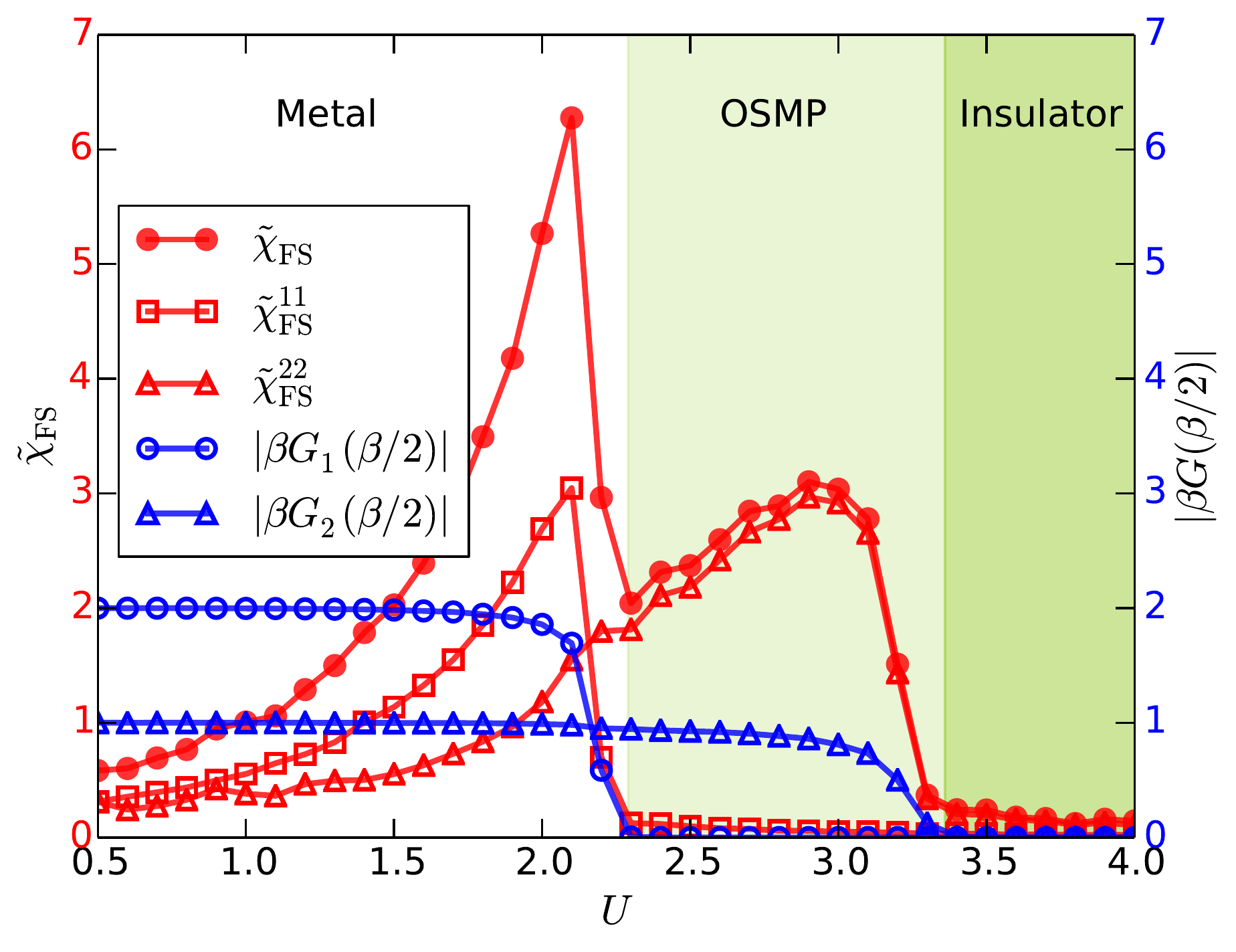}
\caption{(Color online) Orbital-selective Mott metal-insulator transition in the two-band Hubbard model on a Bethe lattice ($\beta = 100.0$, $t_1 = 0.5$, $t_2 = 1.0$, $J = U/4.0$). The abbreviation ``OSMP" means orbital-selective Mott phase. \label{fig:osmt}}
\end{figure}

In multi-orbital Hubbard models, there exist more complicated Mott transition scenarios. For example, if the orbitals are non-degenerate, one may observe a so-called orbital-selective Mott transition (OSMT)~\cite{PhysRevB.85.245110,PhysRevB.79.115119,kunes,PhysRevLett.99.126405}. We will consider such a general two-band Hubbard model with rotationally invariant interaction~\cite{PhysRevLett.92.216402}. The half bandwidths for the two bands are $D_1 = 2t_1 = 1.0$ and $D_2 = 2t_2 = 2.0$, respectively. We calculate the total fidelity susceptibility $\tilde{\chi}_{\text{FS}}$ and orbital-resolved fidelity susceptibility $\tilde{\chi}^{\alpha\gamma}_{\text{FS}}$ as a function of the Coulomb interaction strength $U$ (keeping $J$ fixed to $U/4.0$). The calculated fidelity susceptibilities and the $|\beta G(\beta/2)|$ data are plotted in Fig.~\ref{fig:osmt}.

Based on the calculated $|\beta G(\beta/2)|$ data, $U_c$ is $\sim$ 2.3 for band 1 (narrow band) and $\sim$ 3.4 for band 2 (wide band), which agrees quite well with the previous DMFT + ED (exact diagonalization) results~\cite{PhysRevLett.92.216402}. As for the total fidelity susceptibility, it displays a sharp decline around 2.3, while for $U > 3.4$, it quickly drops to small values.  This behavior can be explained as follows: When $U < 2.3$, the two bands are in a metallic state, and $\tilde{\chi}_{\text{FS}}$ increases with $U$ monotonously. When $U \cong 2.3$, a Mott metal-insulator transition occurs in the narrow band, while the wide band still remains metallic. As a consequence, $\tilde{\chi}_{\text{FS}}$ exhibits the first decline here. At $U \cong 3.4$, the second Mott transition occurs in the wide band. Now the system is in a completely insulating state, and similar to the single-band Hubbard model case, $\tilde{\chi}_{\text{FS}}$ takes small values in this Mott insulator phase. The orbital-resolved fidelity susceptibilities $\tilde{\chi}^{11}_{\text{FS}}$ and $\tilde{\chi}^{22}_{\text{FS}}$ provide a convenient tool to explore the Mott transitions in the two bands individually. They drop to small values at $U \cong 2.3$ and $U \cong 3.4$, respectively, and therefore indicate correctly the positions of the orbital-selective Mott transitions. 

\subsection{High-spin to low-spin transition and Fermi-liquid to non-Fermi-liquid crossover\label{subsec:hsls}}

\begin{figure}[t]
\centering
\includegraphics[width=\columnwidth]{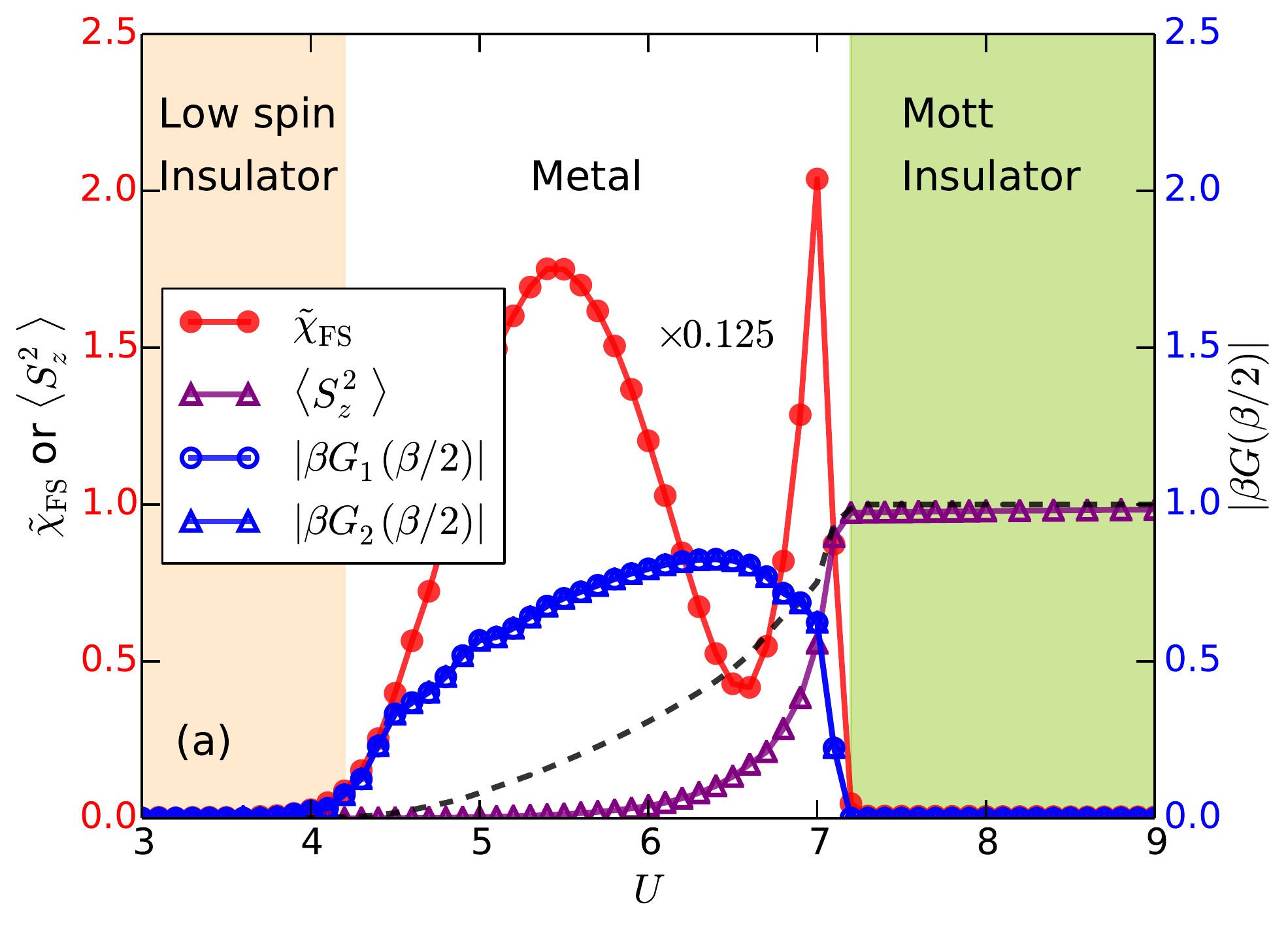}
\includegraphics[width=\columnwidth]{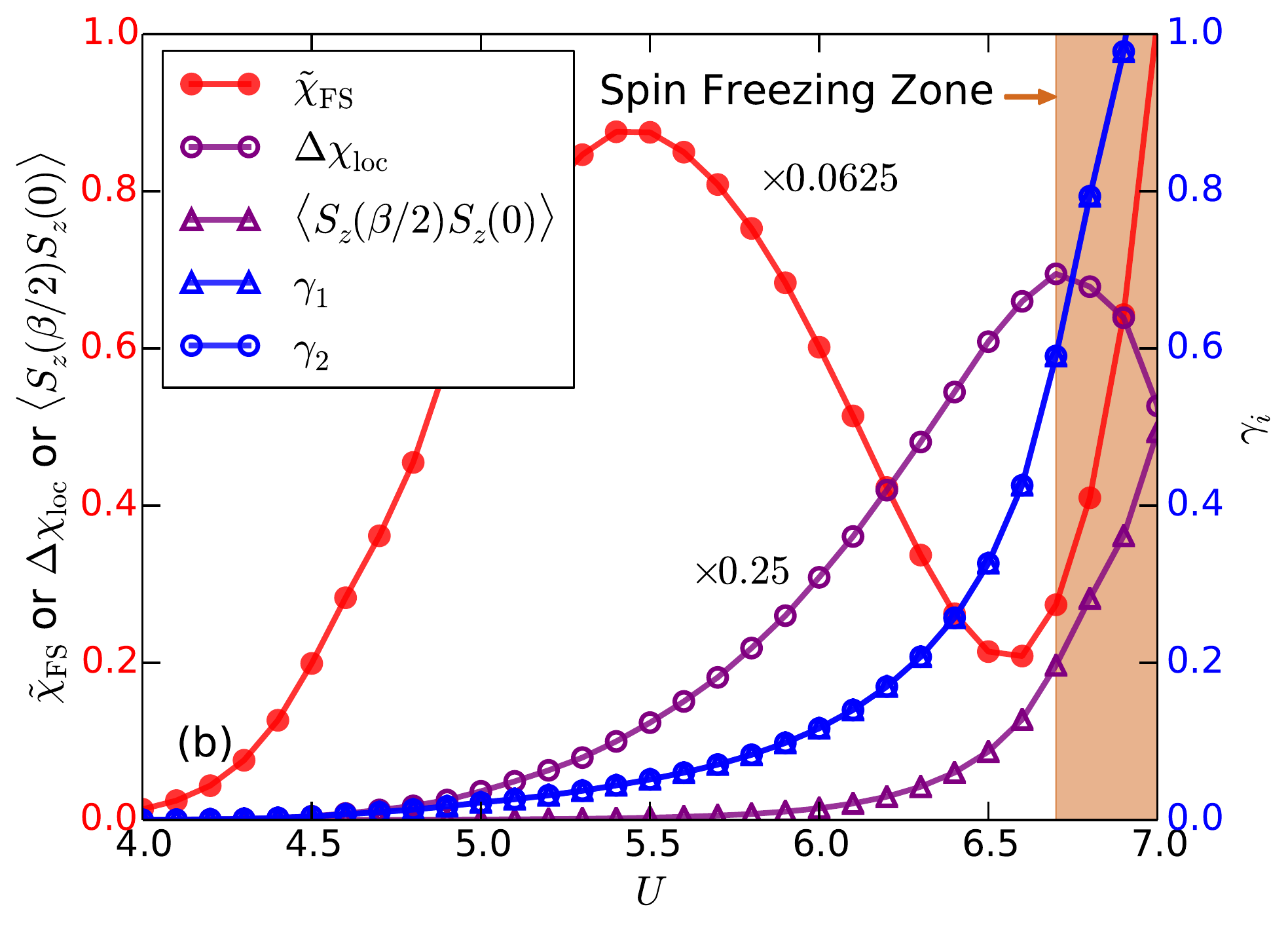}
\caption{(Color online) (a) High-spin to low-spin transition in the two-band Hubbard model on a Bethe lattice ($\beta = 50.0$, $t = 1.0$, $\Delta_{\text{cf}} = 2.5$, $J = U/4.0$). The dashed line shows the occupation number for the high-lying band 1. (b) Fermi-liquid to non-Fermi-liquid and spin-freezing crossovers in the same model. The low-energy scattering rate $\gamma_\alpha$ is defined via Eq.~(\ref{eq:gamma}). The definition for $\Delta \chi_{\text{loc}}$, which can be used to locate the spin-freezing crossover can be found in Eq.~(\ref{eq:cloc}). See main text for more explanations. The $\tilde{\chi}_{\text{FS}}$ and $\Delta \chi_{\text{loc}}$ data are rescaled for a better visualization. \label{fig:hl}}
\end{figure}

We next consider a half-filled two-band Hubbard model with equal bandwidths. Only the density-density type interactions are retained in $\hat{H}_{\text{loc}}$, but an additional crystal-field splitting term $\hat{H}_{\text{cf}} = \Delta_{\text{cf}}\sum_{\sigma} (n_{1\sigma}-n_{2\sigma})$ is included. In the present study, we fix $\Delta_{\text{cf}} = 2.5$~\cite{details} and calculate the total fidelity susceptibility and $\langle S^2_{z}\rangle$ as a function of $U$. The results are shown in Fig.~\ref{fig:hl}(a).

The system switches from a low-spin insulating phase ($\langle S^2_{z}\rangle \sim 0$) to a high-spin Mott insulating phase ($\langle S^2_{z}\rangle \sim 1$) as $U$ is increased. These two distinct insulating phases are separated by a metallic phase in the moderately correlated region~\cite{PhysRevB.85.205106}. In both the high-spin and low-spin insulating regions, $\tilde{\chi}_{\text{FS}}$ is close to zero, while in the metallic phase, $\tilde{\chi}_{\text{FS}}$ becomes relatively large. The critical $U_{c1}$ and $U_{c2}$ for the two  metal-insulator transitions, as determined from $\tilde{\chi}_{\text{FS}}$, agree quite well with the values deduced from the drops in $|\beta G(\beta/2)|$.

In addition, we observe a prominent peak and a deep valley in $\tilde{\chi}_{\text{FS}}$ at $U \cong 5.5$ and 6.5, respectively. The growth of $\tilde{\chi}_{\text{FS}}$ between $U \cong 4.5$ and $5.5$ is likely a signature of increasing correlations and a crossover into a non-Fermi liquid regime. The low-energy scattering rate $\gamma_\alpha$, which is defined as
\begin{equation}
\label{eq:gamma}
\gamma_\alpha = -\Im \Sigma_\alpha(i\omega_n \rightarrow 0),
\end{equation}
can be used to distinguish the Fermi-liquid and non-Fermi-liquid phases~\cite{PhysRevLett.101.166405,PhysRevB.81.054513}. We plot $\gamma_\alpha$ together with the corresponding $\tilde{\chi}_{\text{FS}}$ data of the metallic phase in Fig.~\ref{fig:hl}(b). When $U \leq 4.5$, $\gamma_\alpha$ is close to zero, which is essentially consistent with a Fermi-liquid state. On the contrary, when $U > 4.5$, $\gamma_\alpha$ becomes considerable and grows rapidly, which indicates a non-Fermi-liquid state. On the other hand, the valley near $U = 6.5$ appears to be related to a spin-freezing crossover which competes with the former correlations. Recently, Ref.~[\onlinecite{PhysRevLett.115.247001}] introduced the observable $\Delta \chi_{\text{loc}}$, which measures the local spin fluctuations, to locate the spin-freezing crossover. $\Delta \chi_{\text{loc}}$ is defined as follows:
\begin{equation}
\label{eq:cloc}
\Delta\chi_{\text{loc}} = \chi_{\text{loc}} - \beta \langle S_{z}(\beta/2) S_{z}(0)\rangle,
\end{equation}
where $\chi_{\text{loc}}$ denotes the local magnetic susceptibility~\cite{kunes}:
\begin{equation}
\chi_{\text{loc}} = \int^{\beta}_{0} \langle S_{z}(\tau) S_{z}(0) \rangle d\tau.
\end{equation}
It was suggested that the peak of $\Delta\chi_{\text{loc}}$ can be used to locate the crossover into the spin-frozen regime~\cite{PhysRevLett.115.247001}. Hence, we calculated the spin-spin correlation function $\langle S_{z}(\tau) S_{z}(0)\rangle$ and then used it to extract the $\Delta\chi_{\text{loc}}$. The results are plotted in Fig.~\ref{fig:hl}(b) as well. Since $\langle S_{z}(\beta/2) S_{z}(0)\rangle$ is considerable when $U > 6.0$ and $\Delta \chi_{\text{loc}}$ reaches its maximum value near $U = 6.7$, we conclude that there exists a spin-frozen regime in the metallic phase close to the high-spin Mott insulator [see the color bar in Fig.~\ref{fig:hl}(b)]. This finding is consistent with the very recent results in Ref.~[\onlinecite{hoshino2016}], which demonstrated a mapping between the two-orbital model (with $\Delta_\text{cf}=0$) away from half-filling and the half-filled model with crystal field splitting, which leaves the local moment invariant. In both models, spin-freezing plays an important role and leads to unconventional electronic orders at low temperature.   

\subsection{Spin-freezing crossover\label{subsec:spin}}

\begin{figure*}[t]
\centering
\includegraphics[width=\columnwidth]{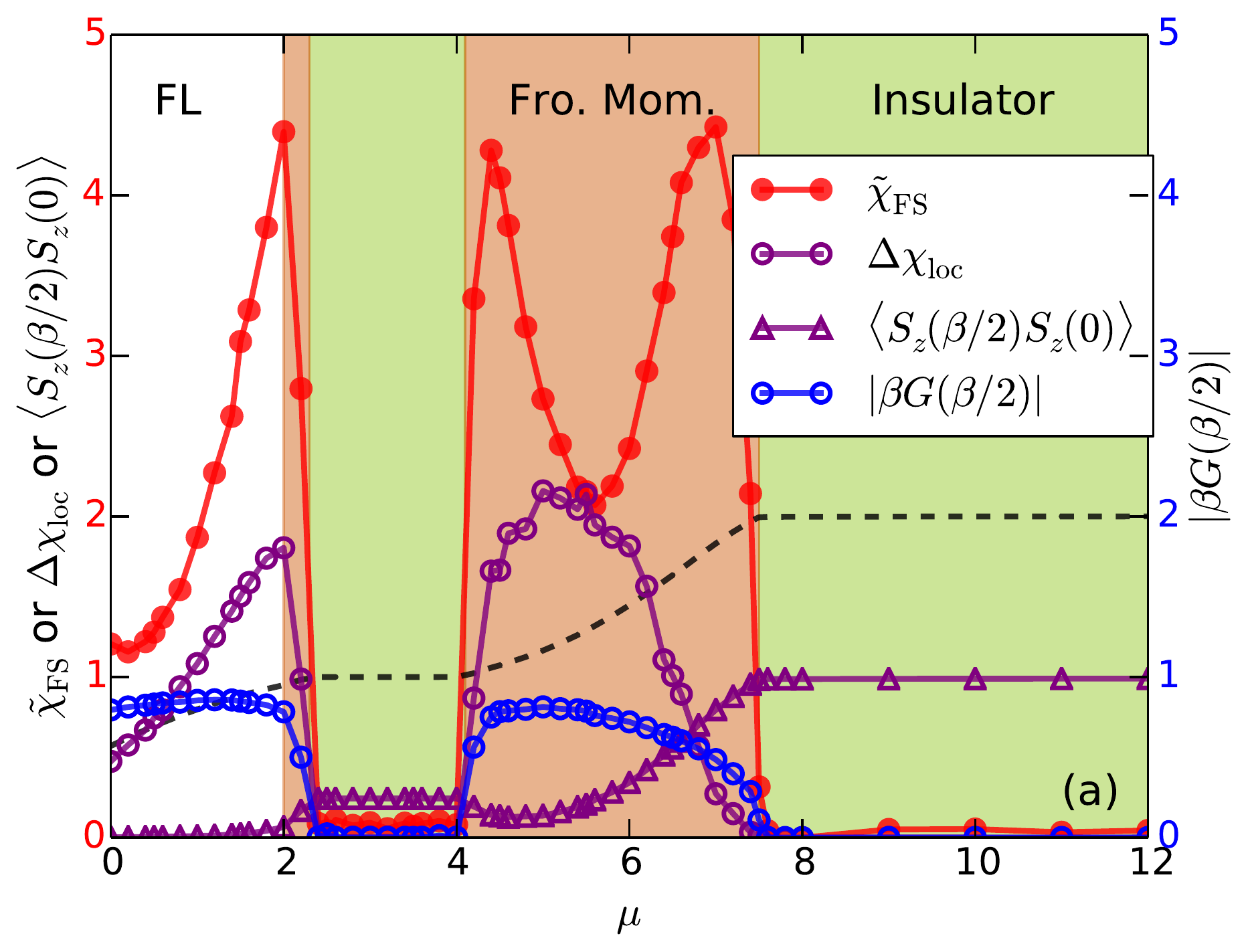}
\includegraphics[width=\columnwidth]{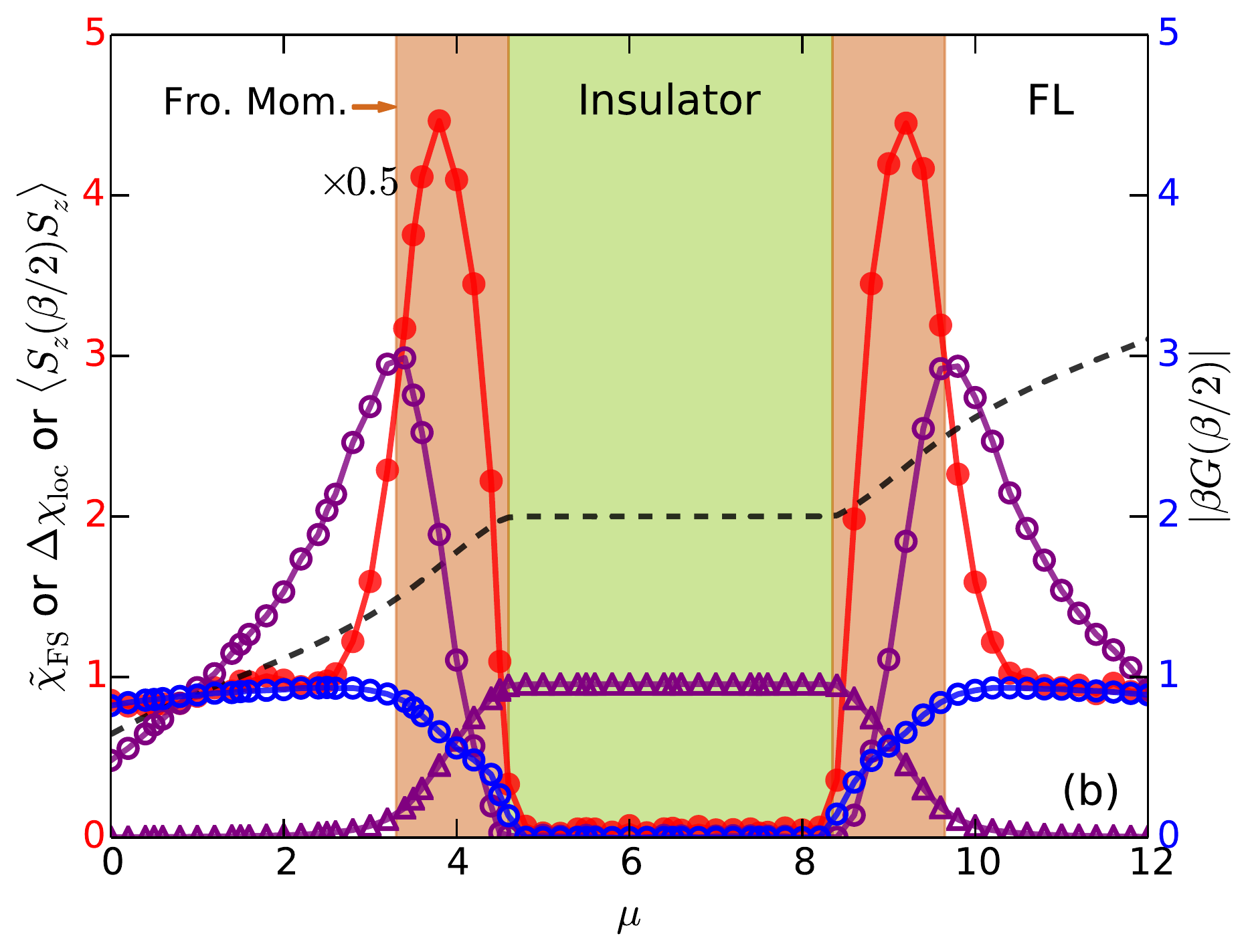}
\includegraphics[width=\columnwidth]{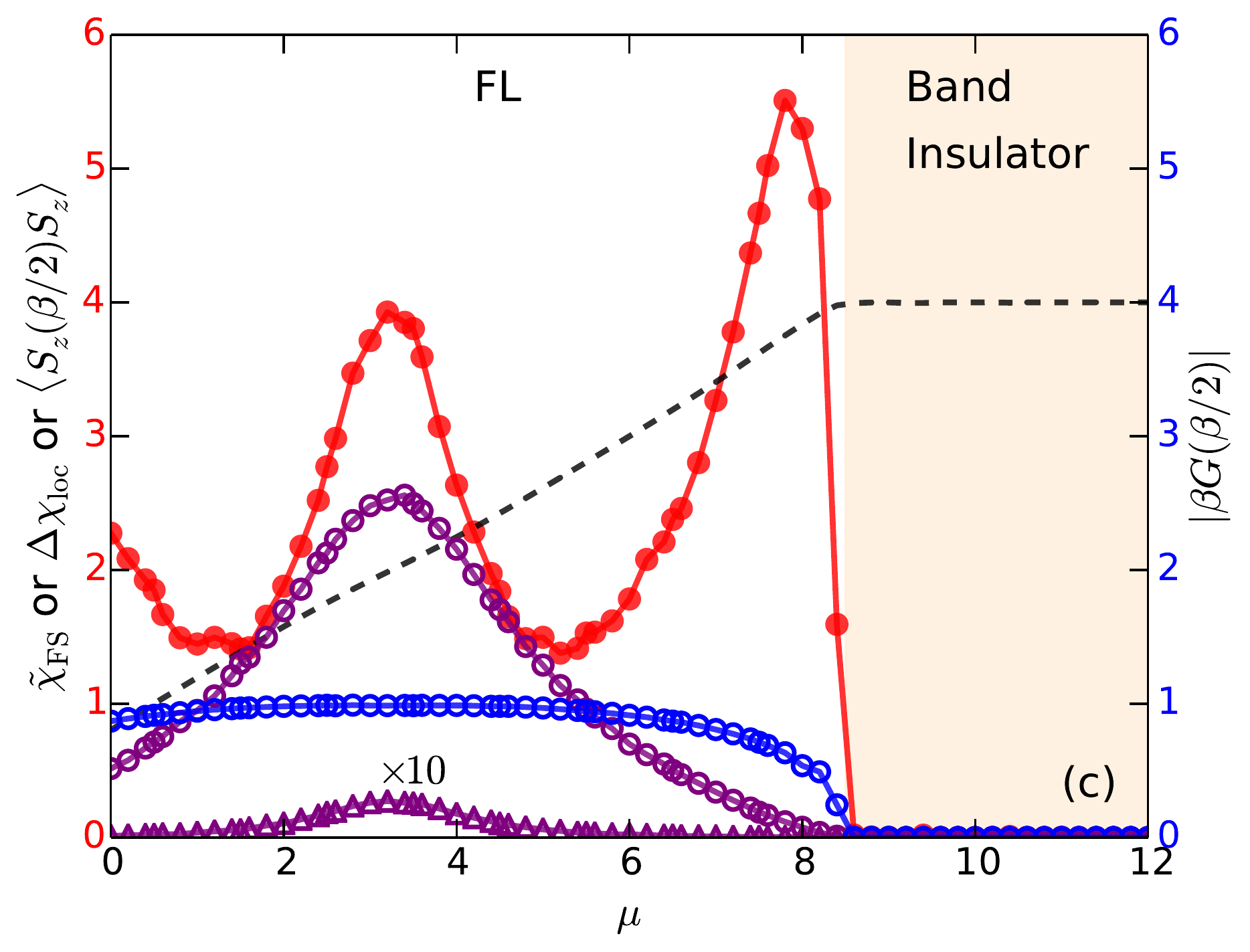}
\caption{(Color online) Spin-freezing crossover in the two-band Hubbard model on a Bethe lattice ($\beta = 50.0$, $t = 1.0$, $J = U/6.0$). (a) $U = 12.0$. (b) $U = 6.0$. (c) $U = 3.0$. The definition for $\Delta \chi_{\text{loc}}$ can be found in Eq.~(\ref{eq:cloc}). The dashed line shows the total occupation number. The abbreviation ``FL" means the Fermi-liquid state, and ``Fro. Mom." the spin-frozen moment phase. In panels (b) and (c), $\tilde{\chi}_{\text{FS}}$ and $\langle S_{z}(\beta/2) S_{z}(0)\rangle$ are rescaled for a better visualization. \label{fig:sf}}
\end{figure*}

Next, we use the generalized fidelity susceptibility to further investigate the crossover into the so-called spin-frozen region in the metallic phase of multi-orbital Hubbard models. Here we study the two-band Hubbard model away from half-filling and assume that the two bands are degenerate. For the sake of simplicity, the spin-flip and pair-hopping terms in $\hat{H}_{\text{loc}}$ are neglected. We consider three different scenarios: (i) strong Coulomb interaction ($U = 12.0$), (ii) intermediate Coulomb interaction ($U = 6.0$), and (iii) weak Coulomb interaction ($U = 3.0$)~\cite{details}. The calculated results are collected and displayed in Fig.~\ref{fig:sf}(a)-(c).

When the Coulomb interaction is strong, this system exhibits a complex sequence of crossovers and phase transitions (Fermi-liquid $\rightarrow $ spin-frozen metallic phase $\rightarrow$ Mott insulator $\rightarrow$ spin-frozen metallic phase $\rightarrow$ Mott insulator) as the chemical potential $\mu$ is increased, as is clearly evident in Fig.~\ref{fig:sf}(a). Let's make a detailed analysis of  these phases, phase transitions and crossovers. First, when $\mu < 2.0$, the total occupation number is less than 1.0 and the system is in a Fermi-liquid metallic state. The fidelity susceptibility $\tilde{\chi}_{\text{FS}}$ increases with increasing $\mu$. Second, for 2.3 $< \mu <$ 4.1 and 7.5 $< \mu <$ 12.0, the total occupation number is very close to 1.0 and 2.0, respectively. In these chemical potential intervals, the system is in a Mott insulating state and $\tilde{\chi}_{\text{FS}}$ is very small, as discussed above. Third, for 4.1 $< \mu <$ 7.5 the system is in the spin-frozen metallic phase, which in the low-temperature regime is characterized by $\langle S_{z}(\tau) S_{z}(0)\rangle$ saturating at long times at a nonzero constant~\cite{PhysRevLett.101.166405,PhysRevB.89.245104}. In this state, $\tilde{\chi}_{\text{FS}}$ is large. The phase boundary between the Mott insulating phase and the spin-frozen metallic phase can be easily identified, since $\tilde{\chi}_{\text{FS}}$ drops rapidly near the critical point. We also notice that $\tilde{\chi}_{\text{FS}}$ shows a ``dip" at $\mu \sim 5.6$. We find some clues to explain it from the behaviors of $\Delta \chi_{\text{loc}}$ and $\langle S_{z}(\beta/2)S_{z}(0)\rangle$. In this filling regime, $\langle S_{z}(\beta/2)S_{z}(0)\rangle$ doesn't increasing monotonously. It decreases at first ($\mu < 5.0$), and then increases ($\mu \geq 5.0$). Correspondingly, $\Delta\chi_{\text{loc}}$ shows a big ``bump" which reaches a maximum at $\mu \sim 5.2$. This behavior indicates that at $\mu \sim 5.0$, the system is in the vicinity of a spin-freezing crossover (from spin-frozen moment phase to Fermi-liquid state), and both by increasing and decreasing the filling, we move deeper into the spin-frozen regime. As a result  $\tilde{\chi}_{\text{FS}}$ and $\langle S_{z}(\beta/2) S_{z}(0) \rangle$ are concave while $\Delta\chi_{\text{loc}}$ is convex as a function of $\mu$. The two-peak structure of $\tilde \chi_\text{FS}$ thus reflects two phenomena: (i) a sharp increase connected to local moment formation as one moves deeper into the spin-frozen regime, and closer to the Mott phases, and (ii) an enhanced fidelity susceptibility in the crossover region with fluctuating local moments. The latter effect prevents an even deeper dip around $\mu\sim 5.0$. Note that this explanation is consistent with the phase diagram obtained in the previous single-site DMFT calculation~\cite{PhysRevB.85.205106}. Fourth, let us discuss the behavior in the metal phase near $\mu=2$. Here, the strong increase of $\Delta\chi_{\text{loc}}$ also suggests the appearance of fluctuating local moments and a proximity to a spin-frozen regime. However, we cannot see a clear maximum in $\Delta\chi_{\text{loc}}$, because the formation of the spin-frozen metal is pre-empted by the Mott transition. Judging from $\langle S_{z}(\beta/2) S_{z}(0) \rangle$, there may be a tiny region with long-lived moments near the $N=1$ Mott insulator, but it is clear that the rapid increase of $\tilde{\chi}_{\text{FS}}$ is primarily driven by the appearance of the still fluctuating local moments.

When the Coulomb interaction is moderate, the situation is a bit different [see Fig.~\ref{fig:sf}(b)]. There exists only an $N=2$ Mott insulating phase in the range 4.6 $ < \mu <$ 8.2, where $\tilde{\chi}_{\text{FS}}$ is small, as discussed above. Around $\mu=3.3$, $\Delta\chi_{\text{loc}}$ reaches a maximum, and the long-time spin-correlation function increases. According to the criterion of Ref.~[\onlinecite{PhysRevLett.115.247001}] the system is thus in the spin-frozen regime for $3.3 \lesssim \mu \lesssim 4.6$. Evidently, the fidelity susceptibility also detects this spin-freezing, as $\tilde{\chi}_{\text{FS}}$ exhibits a rapid increase around $\mu=3.3$. The comparison to $\Delta\chi_{\text{loc}}$ even suggests that $\tilde{\chi}_{\text{FS}}$ provides a more sensitive probe of spin-freezing, with a more narrowly defined crossover region. An analogous behavior is seen on the electron-doped side of the Mott insulating region. 

Finally, when the Coulomb interaction is weak, the phase diagram of this system is less complex [see Fig.~\ref{fig:sf}(c)]. According to the $\tilde{\chi}_{\text{FS}}$ and $|\beta G(\beta/2)|$ data, the system is insulating for $8.5 < \mu < 12.0$ (band insulator with $N=4$). Near filling $N=2$, we observe a single-peak feature in both $\tilde{\chi}_{\text{FS}}$ and $\Delta\chi_{\text{loc}}$. Since $\langle S_{z}(\beta/2) S_{z}(0) \rangle$ is rather small, the system is not in the spin-frozen phase. However, the fluctuations in the local moment are large, which means that the system is  at the verge of a spin-freezing crossover. This conclusion is consistent with Ref.~[\onlinecite{PhysRevB.85.205106}]. If the Coulomb interaction is increased slightly, the system crosses over into a spin-frozen moment phase. The results in Fig.~\ref{fig:sf} show that $\tilde{\chi}_{\text{FS}}$ is very sensitive to the appearance of frozen moments.
 
\section{Discussion and conclusions\label{sec:con}}

As was mentioned in the introduction, the numerically expensive part of a DMFT calculation is the self-consistent solution of a quantum impurity model, which consists of a correlated site coupled to an uncorrelated bath~\cite{RevModPhys.78.865,RevModPhys.68.13}. Many of the phase transitions and crossovers~\cite{PhysRevB.89.245104} occurring in lattice models already manifest themselves at the level of this quantum impurity model. By monitoring the fidelity susceptibility of the auxiliary quantum impurity models one can hence probe crossovers and phase transitions~\cite{PhysRevLett.115.236601,PhysRevX.5.031007} of the original lattice model. 

In a CT-HYB quantum impurity simulation, one naturally measures the fidelity susceptibility defined with respect to the impurity-bath coupling~\cite{PhysRevLett.97.076405,PhysRevB.74.155107,RevModPhys.83.349,PhysRevB.75.155113}. We expect this quantity to be large if the quantum impurity is in the Kondo regime (which corresponds to a Fermi-liquid phase of the lattice model), while it is  small if the impurity is effectively decoupled from its bath (which corresponds to the insulating state of the lattice model). The appearance of local moments also manifests itself at the impurity level~\cite{PhysRevB.89.245104}, and as shown here, the fidelity susceptibility reacts sensitively to the Hund's coupling induced spin-freezing in multi-orbital Hubbard systems.  

In summary, we have calculated the total and orbitally-resolved fidelity susceptibility of single-band and two-band Hubbard models using single-site DMFT with the CT-HYB quantum impurity solver, and found that the fidelity susceptibility can be used to detect various phase transitions and crossovers. The Monte Carlo measurement of the fidelity susceptibility is very cheap and accurate, and the fidelity susceptibility can reveal a phase transition without any priori knowledge about the local order parameter. This makes it an attractive tool for scanning phase diagrams in a systematic and efficient manner. Our work extends and generalizes the application of the fidelity susceptibility to strongly correlated lattice systems. However, there are still many open issues, such as the evolution of the fidelity susceptibility during magnetic phase transitions, its usability in attractive Hubbard models, cluster versions~\cite{RevModPhys.77.1027} and diagrammatic extensions~\cite{PhysRevB.77.033101,PhysRevB.75.045118} of DMFT, etc. Further studies in these directions should be undertaken. 

\begin{acknowledgments}
The Monte Carlo simulations and data analysis have been done using the $i$QIST software package~\cite{iqist}. Most of the calculations were performed on the TianHe-1A machine (in the Tianjin National Supercomputer Center, China). This work was supported by the Natural Science Foundation of China (No.~11504340), the ERC Advanced Grant SIMCOFE, and SNSF Grant No.~200021\_140648.
\end{acknowledgments}

\appendix

\section{More $\kappa$-related statistics}
\renewcommand{\thefigure}{A\arabic{figure}}
\setcounter{figure}{0}

\begin{figure*}[t]
\centering
\includegraphics[width=\columnwidth]{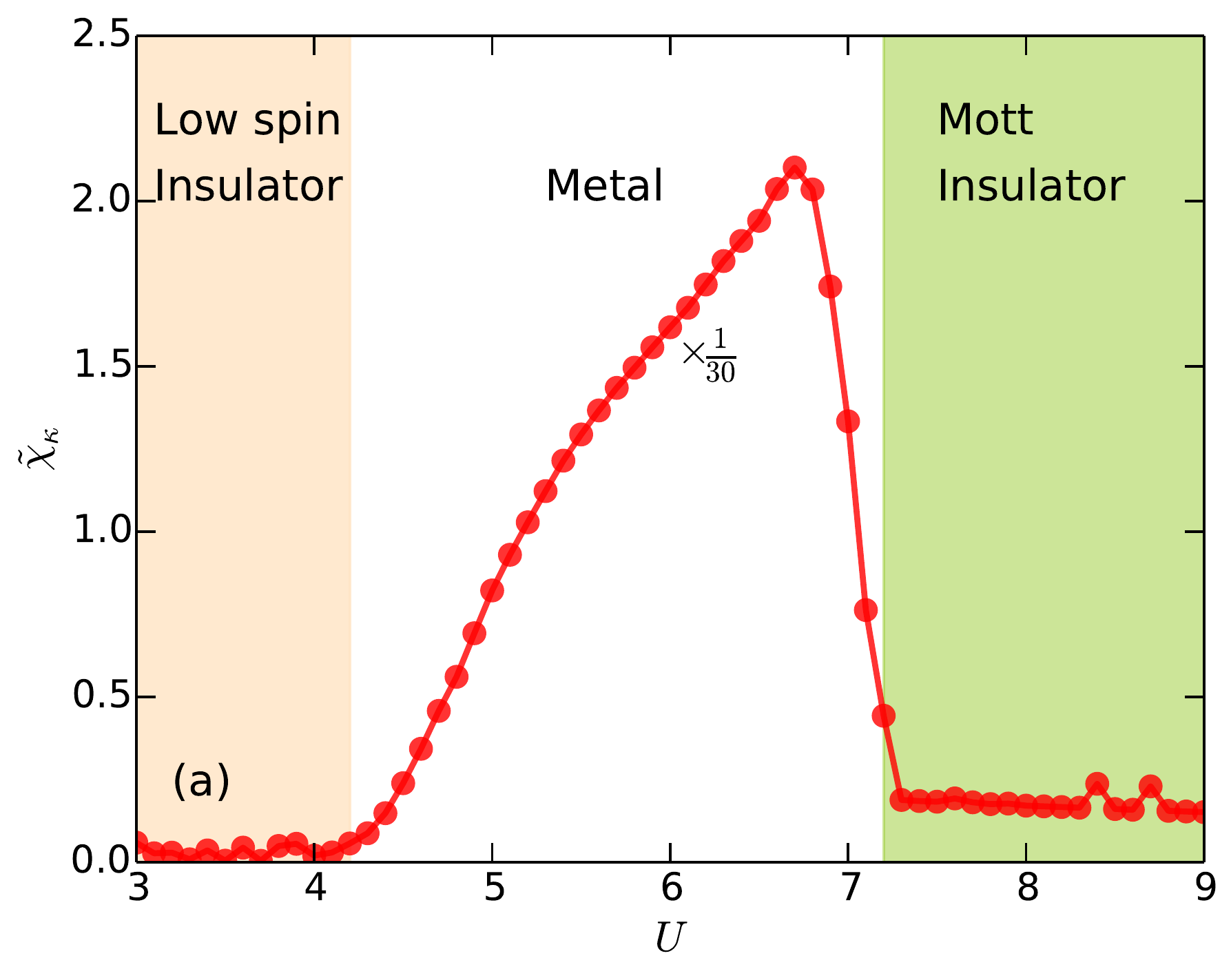}
\includegraphics[width=\columnwidth]{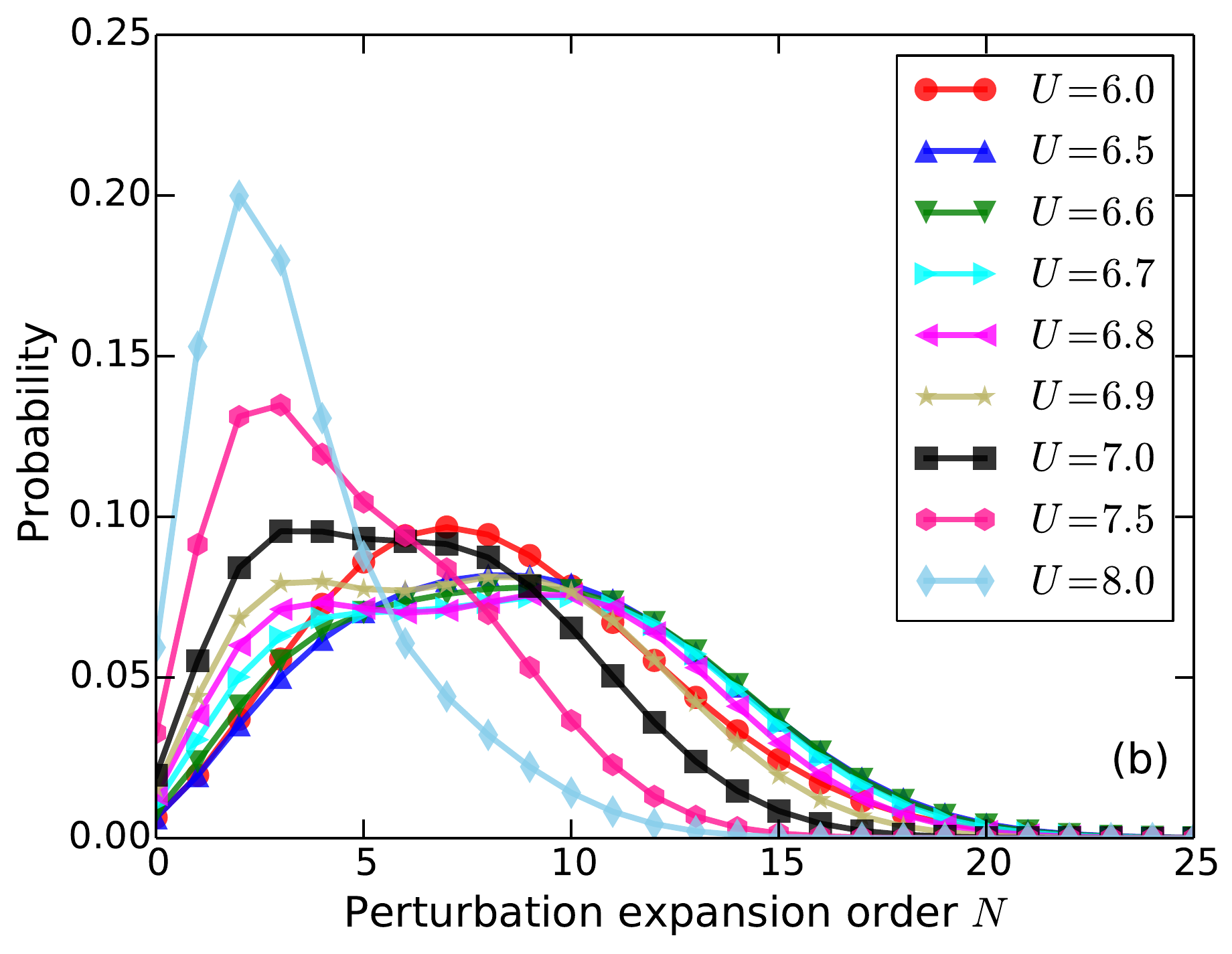}
\includegraphics[width=\columnwidth]{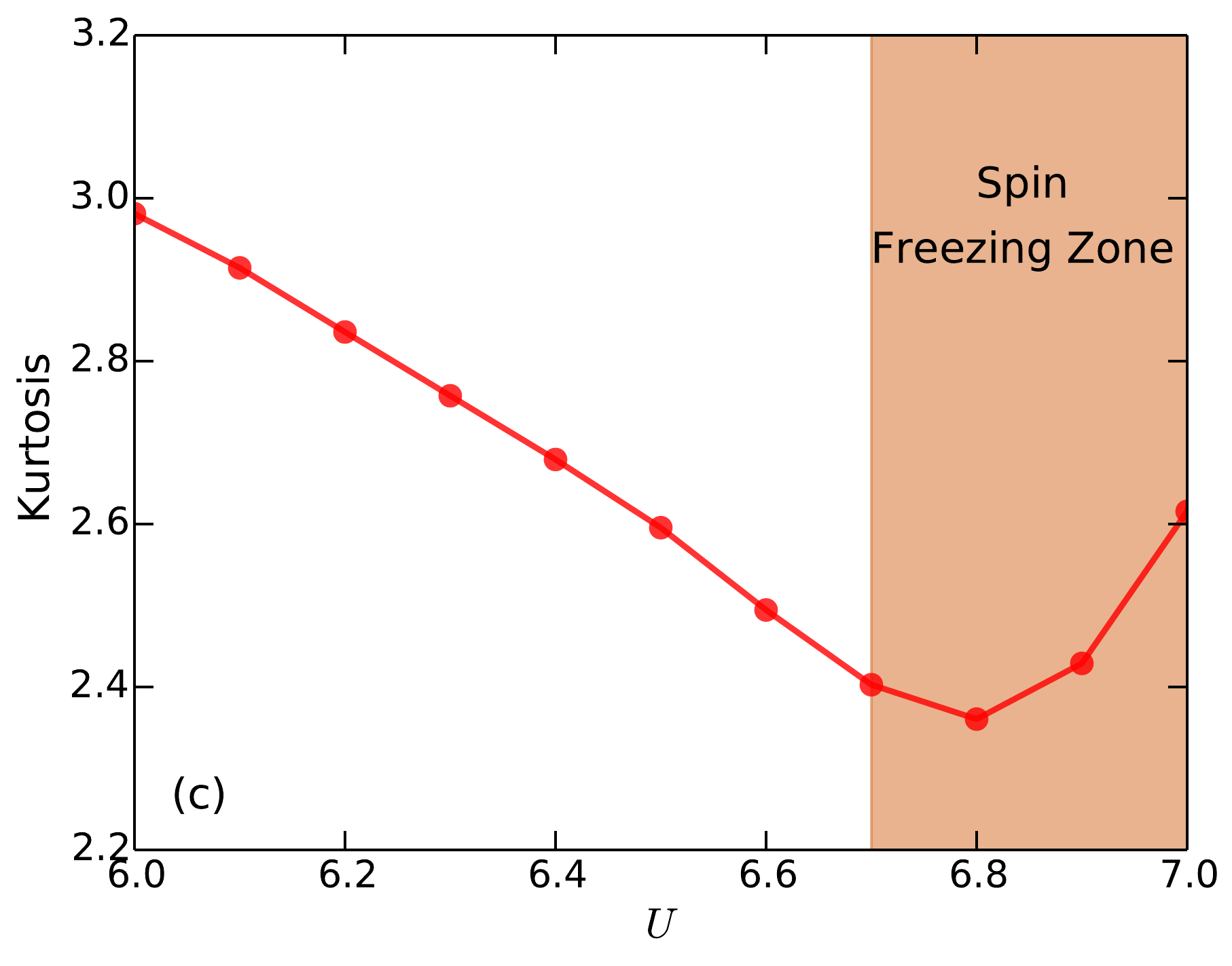}
\caption{(Color online) High-spin to low-spin transition and spin-freezing crossover in the two-band Hubbard model on a Bethe lattice ($\beta = 50.0$, $t = 1.0$, $\Delta_{\text{cf}} = 2.5$, $J = U/4.0$). (a) Generalized variance $\tilde{\chi}_{\kappa}$, which is defined in Eq.~(\ref{eq:kinstd}). The data are rescaled for a better visualization. (b) Histograms of perturbation expansion order $\kappa$ for selected $U$ parameters. (c) Kurtosis analysis of histograms of perturbation expansion order $\kappa$ for selected $U$ parameters. The kurtosis $\gamma_\kappa$ is defined in Eq.~(\ref{eq:kurt}). See text for more explanations. \label{fig:morek}}
\end{figure*}

In the CT-HYB algorithm, the perturbation order $\kappa$ corresponds to the number of hybridization events in a given diagram configuration, and the distribution of perturbation orders is captured by the histogram of $\kappa$, which may be accumulated during the Monte Carlo sampling. Changes in the histogram of $\kappa$ reflect modifications in the impurity-bath coupling, which as mentioned before are tightly connected with the quantum phase transitions and crossovers in the system. The generalized fidelity susceptibility $\tilde{\chi}_{\text{FS}}^{\alpha\gamma}$, which computes the covariance between the number of hybridization events in the left-half and right-half imaginary-time intervals (i.e., $\kappa_{\rm L}$ and $\kappa_{\rm R}$), is very sensitive to the changes in the histogram and the distribution of the $\kappa$, and thus a good quantity to detect various phase transitions and crossovers. 

There are some other $\kappa$-related statistical quantities which allow us to analyze the changes in the histogram, such as the generalized variance ($\tilde{\chi}_{\kappa}$), skewness ($\sigma_\kappa$), and kurtosis ($\gamma_\kappa$) of $\kappa$. Their definitions are as follows:  
\begin{equation}
\label{eq:kinstd}
\tilde{\chi}_{\kappa} = \langle \kappa^ 2 \rangle - \langle \kappa \rangle^2 - \langle \kappa \rangle,
\end{equation}
\begin{equation}
\label{eq:skew}
\sigma_\kappa = \frac{ E[(\kappa - \langle \kappa \rangle)^3] }{E[(\kappa - \langle \kappa \rangle)^2]^{3/2}}, 
\end{equation}
\begin{equation}
\label{eq:kurt}
\gamma_\kappa = \frac{ E[(\kappa - \langle \kappa \rangle)^4] }{E[(\kappa - \langle \kappa \rangle)^2]^{4/2}}. 
\end{equation}
As was already pointed out in Ref.~[\onlinecite{PhysRevX.5.031007}], $\tilde{\chi}_{\kappa}$ is related to the second derivative of the free energy. It resembles the fidelity susceptibility and can be used as an indicator of quantum phase transitions. However, it is easier to locate the critical point using the fidelity susceptibility, since it has a stronger singularity than $\tilde{\chi}_{\kappa}$ near the quantum phase transitions. In this Appendix we will show concrete examples to illustrate the usage of these quantities.

The selected two-band model was already studied in Sec.~\ref{subsec:spin}. Here we calculate $\tilde{\chi}_{\kappa}$, $\sigma_\kappa$, $\gamma_\kappa$ with respect to the Coulomb interaction $U$. The results are shown in Fig.~\ref{fig:morek}. Similar to the $\tilde{\chi}_{\rm FS}$, the generalized variance $\tilde{\chi}_{\kappa}$ becomes very small in the low spin insulator and, to a lesser extent, in the Mott insulator, but exhibits a prominent peak in the metallic region. In other words, it can be used to detect the high-spin to low-spin phase transition. Now let's take a close look at the metallic region. We already know that around $U =$ 4.5 and 6.7, there exists the Fermi-liquid to non-Fermi-liquid and spin-freezing crossovers, which manifest themselves in the double peak structure (centered at $U=$ 5.5 and 6.5) in $\tilde{\chi}_{\rm FS}$. As can be seen in Fig.~\ref{fig:morek}, $\tilde{\chi}_{\kappa}$ shows a ``bump"-like feature at $U=$ 5.5 and a peak at 6.5. This suggests that $\tilde{\chi}_{\kappa}$ is less sensitive to the underlying crossovers than $\tilde{\chi}_{\rm FS}$~\cite{PhysRevX.5.031007}. 

As is clearly seen in Fig.~\ref{fig:morek}(b), when 6.7 $ \leq U \leq$ 7.0, the histograms of perturbation expansion orders $\kappa$ show an unusual double-peak structure, with the lower (higher) peak resembling the histogram of the Mott insulating (metallic) solution. This structure, which appears in the spin-frozen regime, is a signature for the emergence of local moments. As a result of the double peaks, the histograms deviate strongly from the normal distribution. Such deviations can be quantified by the skewness and kurtosis. In Fig.~\ref{fig:morek}(c), we observe a minimum of the kurtosis $\gamma_\kappa$ near $U=$ 6.8, which is very close to the critical point for the spin-freezing crossover ($U \sim$\ 6.7) determined by the peak position of $\Delta\chi_{\text{loc}}$~\cite{PhysRevLett.115.247001}. The skewness $\sigma_\kappa$ also displays a minimum at the same position (not shown in this figure). Therefore, $\gamma_\kappa$ and $\sigma_\kappa$ are also promising tools for detecting a spin-freezing crossover. Since their calculations only involve the accumulation of statistics for hybridization events, they should be more effective than the numerically more demanding calculation of $\Delta\chi_{\text{loc}}$.   


\bibliography{fidel}

\begin{thebibliography}{39}
\bibitem [{det()}]{details}%
  \BibitemOpen
  \href@noop {} {}\bibinfo {note} {This setup is the same as in previous DMFT
  calculations. See Ref.~\onlinecite{PhysRevB.85.205106}.}\BibitemShut {Stop}%
\end{thebibliography}

\end{document}